\documentclass[twocolumn,showpacs,amsmath,amstex,amssymb,mathfonts,prl]{revtex4}
\usepackage{graphicx,bm,units}

\usepackage{color}

\begin{document}

\newcommand{\vk}{{\bf k}}
\def\ns{^{\vphantom{*}}}
\def\ket#1{{|  #1 \rangle}}
\def\bra#1{{\langle #1|}}
\def\braket#1#2{{\langle #1  |   #2 \rangle}}
\def\expect#1#2#3{{\langle #1 |   #2  |  #3 \rangle}}
\def\cH{{\cal H}}
\def\half{\frac{1}{2}}
\def\sut{\textsf{SU}(2)}
\def\suto{\textsf{SU}(2)\ns_1}
\def\kF{\ket{\,{\rm F}\,}}

\title{Tunable Electron Interactions and Fractional Quantum Hall
  States in Graphene}
\author{Z. Papi\'c$^1$, R. Thomale$^2$, and D. A. Abanin$^{2,3}$}
\affiliation{$^1$ Department of Electrical Engineering, Princeton University, Princeton, NJ 08544}
\affiliation{$^2$ Department of Physics, Princeton University, Princeton, NJ 08544}
\affiliation{$^3$ Princeton
  Center for Theoretical Science, Princeton University, Princeton, NJ
  08544}

\date{\today}

\begin{abstract}
The recent discovery of fractional quantum Hall states in graphene raises
the question of whether the physics of graphene offers any advantages over GaAs-based materials in exploring strongly-correlated states of two-dimensional electrons. Here we propose a method to continuously tune the effective electron interactions in graphene and its bilayer by the dielectric environment of the sample. Using this method, the charge gaps of prominent FQH states, including $\nu=1/3$ or $\nu=5/2$ states, can be increased several times, or reduced to zero. The tunability of the interactions can be used to realize and stabilize various strongly correlated phases and explore the transitions between them.

\end{abstract}


\date{\today}

\maketitle

{\sl Introduction.} Two-dimensional electron systems (2DES) placed in
a high magnetic field exhibit strongly correlated phases
characterized by fractionally quantized Hall conductivity~\cite{tsg},
quasiparticles that carry a fraction of electron charge~\cite{laughlin}, and
fractional (Abelian or possibly non-Abelian) statistics~\cite{laughlin, mr}.
These remarkable phenomena occur in the extreme quantum limit -- the fractional quantum Hall (FQH) regime --
when the number of electrons, $N_e$, is comparable to the number of magnetic flux quanta through the 2DES, $N_\Phi$,
corresponding to a partial filling $\nu=N_e/N_\Phi$ of one of the lower Landau levels (LLs).
When $\nu$ is swept through the series of simple fractions in the lowest $n=0$ Landau level (LLL) where $N_e<N_\Phi$, the electrons  
condense into Laughlin states which describe the strongest observed fractions ($\nu=1/3,1/5,\ldots$)~\cite{laughlin}, or weaker states that belong to the so-called hierarchy~\cite{prange, halperin}
or composite fermion series~\cite{jainbook}. 
Within the $n=1$ LL, the effective interaction changes due to the nodal structure of LL orbitals, and some of the more exotic states, such as the Read-Rezayi states~\cite{rr_parafermion}, are more likely to be favored. The experimentally most important member of the Read-Rezayi series is the
Moore-Read (MR) ``Pfaffian" state~\cite{mr}, believed to describe the FQH plateau at $\nu=5/2$~\cite{eover4, gww}. Quasiparticles of the Read-Rezayi series obey the non-Abelian statistics~\cite{mr} which is of interest for topological quantum computation~\cite{tqc}.

Although many FQH states have been discovered in GaAs-based 2DES, these systems are plagued by the fact that their 2DES is buried inside a larger 3D structure. This fixes the effective interactions at values that are often not optimal for some of the most interesting FQH states, including the Read-Rezayi series. Theoretically, such states are known to be very sensitive to the form of the effective interactions~\cite{morf_pf,rh_pf}.
Another problem stems from the strong dielectric screening and finite well-width~\cite{zds} in GaAs, which weaken the electron-electron interactions, thereby making FQH states fragile. This has been a major obstacle in the studies of the possibly non-Abelian states, which could only be observed in ultra-high-mobility samples~\cite{eover4} (see however~\cite{muraki}). Thus, it is desirable to find an alternative high-mobility 2DES with strong effective Coulomb interactions that are adjustable in a broad range.

A promising candidate for this kind of material is monolayer graphene (MG), a high-mobility atomically thick 2DES~\cite{castroneto}, where recently a $\nu=1/3$ state~\cite{graphene_fqhe}, and several other states~\cite{dean}, have been discovered. Remarkably, due to graphene's truly 2D nature, the short-range electron interactions greatly exceed those in GaAs, which leads to a significantly more robust FQH state at $\nu=1/3$~\cite{abanin-10prb115410}. A closely
related material, bilayer graphene (BG)~\cite{castroneto}
has similarly high mobility, and exhibits interaction-induced quantum Hall states at integer filling factors at low magnetic fields~\cite{feldman-09np889}. This indicates that, similarly to MG,
the underlying electron interactions are strong and one could expect robust FQH states in BG as well.

\begin{figure}[ttt]
\vspace{-1mm}
   \includegraphics[scale=0.6]{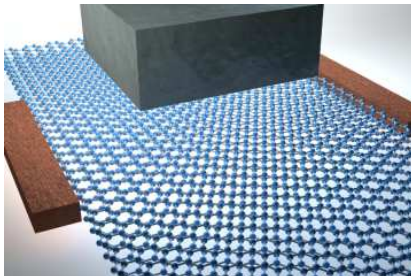}
  \caption{(Color online). Setup of a graphene system with tunable electron
    interactions. Electron states are affected by a dielectric
    plate placed in the vicinity of the surface. By varying the dielectric 
     permittivity of the plate and its distance $d/2$ to the graphere layer,
   interactions of the desired form can be engineered.}
\label{fig1}
\end{figure}

{\sl Main results.}  Here we propose a way to continuously tune the interactions in graphene in a wide range, using the fact that 2DES in graphene is exposed and its properties can be controlled by the dielectric environment, as illustrated in Fig.~\ref{fig1}. Using exact diagonalization calculations, we show that the tunability can be used to significantly increase the excitation gap of $\nu=1/3$ state in both MG and BG, as well as in the half-filled $n=1$ LL in BG. The latter, similarly to the case of GaAs, resides in the gapped MR Pfaffian state~\cite{morf_pf, rh_pf}. As the gap is varied, the overlaps between the exact states and the model
wavefunctions improve by a few percent, and their topological character becomes better protected. The reduction of the gap induces the transitions to compressible and crystalline phases~\cite{rh_pf}.

{\sl Model.} A graphene sample is situated in a dielectric medium with permittivity $\epsilon_1$, and a semi-infinite dielectric plate with permittivity $\epsilon_2\neq \epsilon_1$ is placed at a distance
$d/2$ away from the graphene sheet. The effective interactions between
electrons in graphene change due to the surface charges induced at the boundary
between dielectrics:
\begin{equation}\label{eq:interaction_screened}
V(r)=\frac{e^2}{\epsilon_1 r}+\alpha \frac{e^2}{\epsilon_1\sqrt{r^2+d^2}}, \,\,
\alpha=\frac{\epsilon_1-\epsilon_2}{\epsilon_1+\epsilon_2}.
\end{equation}
Below we measure the distance $d$ in units of the magnetic length $l_B$ and the energy in units of $e^2/\epsilon_1 l_B$. Ratio $d/l_B$ controls the effective interactions within a partially filled LL (see below). 
All the gaps quoted here should be multiplied by a factor $\epsilon_{\rm GaAs}/\epsilon_1$ if comparison is to be made with GaAs 2DES. An important advantage of this setup is that the interactions can be tuned {\it in situ} by varying the magnetic field $B$, which modifies the ratio $d/l_B$.

We project the interaction to a partially filled LL under consideration, following the standard procedure in FQH numerical studies~\cite{ed}. Within a LL, $V(r)$ is parameterized by the Haldane pseudopotentials $\{ V_m \}$~\cite{ed}, which can be conveniently evaluated from the Fourier transform of the Coulomb interaction, $\tilde{V}(\mathbf{q})$, and the form factor $F(\mathbf{q})$ encoding the properties of LL orbitals. Due to the difference in band structure, the form factors in MG, BG are generally distinct from those of GaAs~\cite{goerbig-06prb161407}. 

For what follows, it is useful to review the previous results on the relation between pseudopotential values and the stability of $\nu=1/3$ and $\nu=5/2$ states.
At $\nu=1/3$, the bare Coulomb interaction in $n=0$ LL favors the Laughlin state. Reducing $V_1$ while keeping $V_{m\geq 3}$ constant eventually destroys the
gap and a compressible state sets in~\cite{ed}. At half filling of the LLL ($\nu=1/2$), the Coulomb interaction gives rise to a gapless Fermi liquid of composite fermions~\cite{hlr}. However, in half-filled $n=1$ LL there is a fully developed plateau in experiments~\cite{eover4}, attributed to the MR state~\cite{mr}. MR state is an eigenstate of a particular three-body repulsive interaction~\cite{gww}. 
Remarkably, in numerical studies, the ground state of the Coulomb interaction at $\nu=5/2$ is seen to be adiabatically connected to the MR state~\cite{storni, thomale}; the overlap of the ground state with the MR state is improved by
the increase in $V_1$ pseudopotential (or, alternatively, by reducing $V_3$)~\cite{rh_pf}. 
Therefore, theory shows that varying the first few pseudopotentials provides
a convenient way to assess stability and induce transitions between FQH states. However, so far it has been difficult to find a controlled way of tuning $V_m$'s experimentally in a sufficiently broad range.

\begin{figure}[ttt]
\vspace{-1mm}
\includegraphics[scale=0.4]{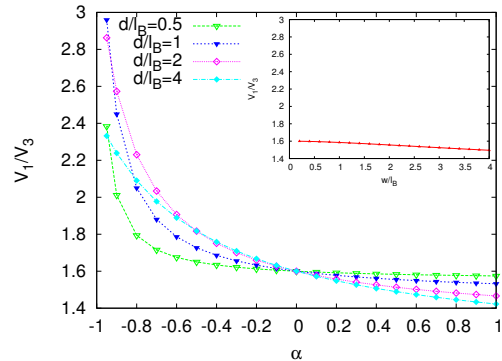}
\caption[]{(Color online). Ratio $V_1/V_3$ as a function of $\alpha$ for several values of $d/l_B$ in $n=0$ LL of MG and BG. 
Inset: typical ratio $V_1/V_3$ for a GaAs infinite quantum well of width $w/l_B$, plotted on the same scale.}
\label{fig2}
\end{figure}
In Fig.~\ref{fig2} we show an example of the ratio $V_1/V_3$ calculated for $n=0$ LL of MG and BG using the setup illustrated in Fig.~\ref{fig1}. Changing $\alpha$ and $d/l_B$ leads to a large variation of $V_1/V_3$ with respect to its pristine value ($\alpha=0$). For comparison, we also show a typical variation of $V_1/V_3$ achievable in GaAs wide quantum wells (inset). 
In what follows, we investigate the model defined by Eq. (\ref{eq:interaction_screened}) using exact diagonalization studies~\cite{ed}. We focus on the case of
spin- and valley-polarized FQH states, such as the Laughlin or MR state, and consider the physics of a single non-degenerate Landau sub-level. For example, in case of the MR state, we expect the total filling factor $\nu=-4+2M+3/2$ of the BG to be the most suitable for realizing the Pfaffian state. Note that the exchange interactions favor the following order of filling the eight $n=0,1$ Landau
sub-levels of the BG~\cite{bilayer_ff}: $ |0\sigma _1 s_1 \rangle $, $|1\sigma_1 s_1\rangle$, $|0\sigma_2 s_2\rangle$, $|1\sigma_2 s_2\rangle$, where $|\sigma_{i} s_i\rangle $, $1\leq i\leq 4$ are the four orthogonal states in the space of
internal indices, with $s$ and $\sigma$ denoting two spin and two valley species, respectively. At $\nu=-4+2M+3/2$, 
where $M=0,1,2,3$, the $M$ pairs of $n=0,1$ sub-levels are expected to be filled, and in the top-most pair, 
$n=0$ Landau sub-level is filled completely, as dictated by the exchange interactions, while $n=1$ sub-level is
half-filled. For brevity, below we refer to this $n=1$ sub-level as ``$n=1$ LL in BG".

\begin{figure}[ttt]
\vspace{-1mm}
\includegraphics[scale=0.5]{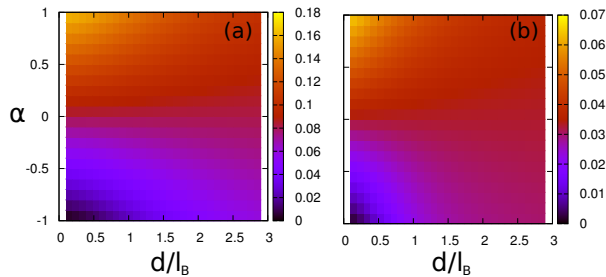}
\caption[]{(Color online). Charge gaps for $N=10$ particle $\nu=1/3$ state in the $n=0$ LL of MG and BG (a), and $N=14$ particle $\nu=1/2$ filled $n=1$ LL of BG (b). By changing $\alpha$ and $d/l_B$, the gaps can be increased several times or brought down to zero, following similar trends in each of the LLs.}
\label{fig3}
\end{figure}
One important criterion of the stability of FQH states is their gap to excitations, which can be either neutral (quasiparticle-quasihole pair above the ground state) or charged (created by changing the magnetic flux by one unit). These two gaps are generally different, but for the model defined by Eq. (\ref{eq:interaction_screened}) we find that they show the same behavior as a function of parameters $\alpha$ and $d/l_B$. We therefore focus on the charge gap, defined by $\Delta_c\equiv E_{qh}+E_{qp}-2E_0$, where $E_0$ is the ground state energy and $E_{qh},E_{qp}$ are the energies at $N_\Phi+1,N_\Phi-1$, respectively ($N_e$ is kept fixed). In evaluating $\Delta_c$, it is convenient to place the FQH system on the surface of a sphere, where incompressible ground states are easily identified by their zero angular momentum. On a finite sphere, FQH states are further characterized by the shift $\mathcal{S}$~\cite{shift}, which relates $N_e,N_\Phi$ to the filling factor $\nu$ in the thermodynamic limit via $N_\Phi=N_e/\nu+\mathcal{S}$. In Fig.~\ref{fig3} we plot $\Delta_c$ for the case of $\nu=1/3$ filling in the $n=0$ LL (MG, BG) and $\nu=1/2$ filling of $n=1$ LL in BG. We do not show the results for the $n=1$ LL of MG because they support significantly weaker pairing correlations than in BG~\cite{graphene_cfl}.
In these plots, the system size is fixed at $N=10$ ($\nu=1/3$) and $N=14$ ($\nu=1/2$), and $\mathcal{S}$ is chosen to be $-3$. This shift corresponds to the case of bare Coulomb interaction where the states are known to be described by Laughlin and MR wavefunctions. We estimate that the gaps can be increased $2-3$ times with respect to the vacuum value ($\alpha=0$). The gaps in $n=1$ LL of BG are overall significantly smaller than those in $n=0$ LL, which is consistent with the experiments detecting fewer FQH states in $n=1$ LL of GaAs. When $\alpha<0$, the gaps become very small and our analysis would likely need to be extended to other shifts that may be preferred by these compressible states in minimizing their ground-state energy. This analysis will be provided elsewhere. We emphasize that all the systems smaller than those in Fig.~\ref{fig3} show qualitatively identical plots with the same relative variation of the gap (we use the rescaled magnetic length and background charge correction in Fig.~\ref{fig3} in order to minimize the finite-size effects~\cite{jainbook}).

Comparing the gaps in Fig.~\ref{fig3} with the ratio $V_1/V_3$, we note that the maximum value of $V_1/V_3$ does not coincide with the maximum of the gap. This is because the gaps are determined by the complete set of pseudopotentials -- while negative $\alpha$'s increase the ratio $V_1/V_3$, they also reduce the magnitude of each individual $V_m$, thereby lowering the gap. However, the Laughlin states and higher order hierarchy states are excellent trial states for positive, as well as negative, $\alpha$ i.e. for long-range Coulomb-like interaction as well as strong short-ranged repulsion. This is reflected in the high value of the overlap between an exact state and the Laughlin wavefunction which varies by as little as $\pm 1\%$ over the entire phase diagram in Fig.~\ref{fig3} (not shown).

\begin{figure}[ttt]
\vspace{-1mm}
\includegraphics[scale=0.45]{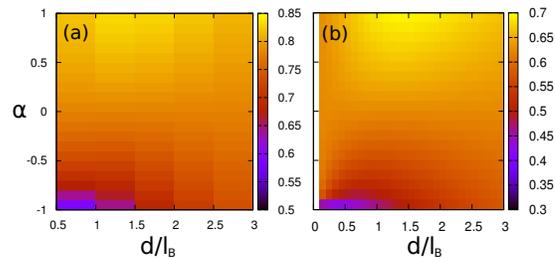}
\caption[]{(Color online). Overlaps between the exact ground state, for each $\alpha$ and $d/l_B$, and the MR Pfaffian state, for $N=16$ particles on the sphere (a), and $N=14$ particles on the torus (b). Maximum overlap occurs for intermediate $d/l_B$.}
\label{fig4}
\end{figure}
The stability of the MR Pfaffian state is more subtle and we expect sharper variation of the overlap as a function of $\alpha$ and $d/l_B$.
In Fig.~\ref{fig4} we show the overlap between the exact state of the dielectric model (for given $\alpha$ and $d/l_B$) and the MR Pfaffian state. The overlaps are remarkably consistent for the two choices of boundary conditions: on the sphere with $\mathcal{S}=-3$, Fig.~\ref{fig4}(a), and on the torus with the hexagonal unit cell, Fig.~\ref{fig4}(b). Although MR state is threefold degenerate and characterized by different pseudomomenta $\mathbf{k}_i;i=1,2,3$ on the torus~\cite{rh_pf}, the sixfold symmetry of the Bravais lattice guarantees the degeneracy of these $\mathbf{k}_i$ subspaces and makes it possible to calculate the overlap in the usual way. The maximum of the overlap occurs for $d$ between $l_B$ and $2l_B$, and is somewhat higher on the sphere, due to the bias of the shift which lifts the particle-hole symmetry (the anti-Pfaffian~\cite{antipf}, having $\mathcal{S}=1$, occurs in a different Hilbert space and does not ``mix" with the Pfaffian, like on the torus). 

\begin{figure}[ttt]
\includegraphics[width=\linewidth]{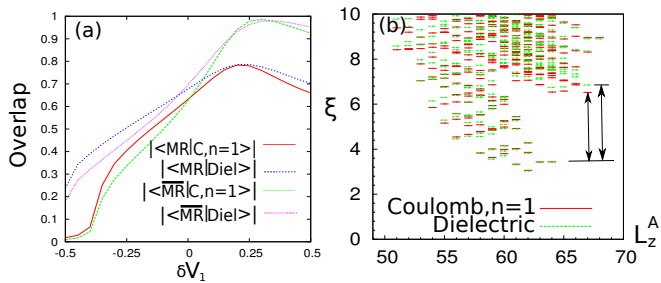}
\vspace{-1mm}
\caption[]{(Color online). (a) Overlaps between the (un)screened Coulomb interaction eigenstate and MR Pfaffian, as well as particle-hole symmetrized MR, for $N=14$ particles on the torus. Screened Coulomb eigenstate is defined by $\alpha=1,d=1.5l_B$ (compare with Fig.~\ref{fig4}(b)). (b) Entanglement spectrum on the sphere for $N=16$ electrons, for bare and screened Coulomb interaction. Entanglement cut is chosen as $0|1$ in the notation of Ref.~\cite{lihaldane}; other cuts yield similar results.}
\label{fig5}
\end{figure}

We now present additional evidence in favor of the state being the MR Pfaffian (Fig.~\ref{fig5}). First, we track the evolution of the state upon \emph{further} variation of $V_1$ pseudopotential, Fig.~\ref{fig5}(a). The increase of $V_1$ would partly mimic the effects of LL mixing~\cite{llmixing} that we neglected so far (we do not investigate the effect of three-body interaction that also arises in this case). In Fig.~\ref{fig5}(a) we show the overlaps between the exact state, $|{\rm Diel}\rangle$, at $\alpha=1,d=1.5l_B$ (chosen from the high-overlap region in Fig.~\ref{fig4}(b)), the bare Coulomb $n=1$ LL state, $|{\rm C,n=1}\rangle$, with the MR Pfaffian $|{\rm MR}\rangle$ and its particle-hole symmetrized version, $|{\overline{\rm MR}}\rangle$. The bare Coulomb state and the dielectric state behave very similarly under the change of $V_1$, the latter being somewhat more robust. As noted in Ref.~\cite{rh_pf}, the overlaps with $|{\overline{\rm MR}}\rangle$ on the torus are large ($97\%$ and better), which is again the case for our dielectric state as well.

Another independent insight into the nature of the dielectric state is the entanglement spectrum~\cite{lihaldane}. The multiplicities of the low-lying levels of the entanglement spectrum contain topological information about a FQH state; this information is protected by an entanglement gap~\cite{thomale} whose magnitude measures the topological stability of a state.  
We calculate and compare the entanglement spectra on the sphere for the $\nu=1/2$ Coulomb $n=1$ LL state and the dielectric eigenstate tuned to a region of high overlap with the MR Pfaffian (Fig.~\ref{fig5}(b)). Not only do the both spectra display the same counting as the MR state (up to some value of the subsystem $A$'s projection of the angular momentum $L_z^A$, set by the finite size of the sphere), but the entanglement gap (indicated by arrows in Fig.~\ref{fig5}(b)) is also somewhat enhanced with respect to the unscreened Coulomb interaction. 

{\sl Conclusions.} In summary, we proposed a method to tune the electron interactions in graphene and its bilayer. In this approach, interaction pseudopotentials can be varied in a broad interval and FQH gaps can be enhanced several times or even reduced to zero, allowing for a more complete exploration of compressible and incompressible phases than can be attained in GaAs materials. The proposed method is expected to be very efficient in optimizing the Abelian FQH states that belong to the hierarchy series, where the variation of the charge gap is followed by a small change in the overlap that generally remains very close to unity. Non-Abelian states, although expected to be stabilized, may require a more subtle approach with several dielectric plates of different permittivities and thicknesses of the order of $l_B$ placed in the vicinity of the surface. In such a system, interactions can be tuned in a broader range and would admit simultaneous change of several pseudopotentials that may be required for the realization of other non-Abelian and multi-component FQH states~\cite{papic_graphene}. Finally, we note that the experiments on the tunneling density of states~\cite{dial} can directly measure the values of the tuned pseudopotentials~\cite{macdonald10}.

{\sl Acknowledgements}. We thank E. Andrei, C. Dean, P. Kim, C. Nayak, Y. Zhang, M. Goerbig, R. Bhatt and especially S. Sondhi for insightful discussions. This work was supported by DOE grant DE-SC$0002140$. DA thanks Aspen Center for Physics for hospitality. RT is supported by a Feodor Lynen fellowship of the Humboldt Foundation.

\bibliography{paper}

\end{document}